\documentclass[preprint,showpacs,preprintnumbers,amsmath,amssymb]{revtex4-1}
\usepackage{graphics,epsfig,subfigure}
\usepackage{diagbox}
\usepackage[usenames]{color}
\usepackage[colorlinks,
linkcolor=blue,
anchorcolor=red,
citecolor=red
]{hyperref}
\usepackage{color}
\usepackage{graphicx}
\usepackage{amsmath}
\usepackage{amsfonts}
\usepackage{amssymb}
\usepackage{txfonts}
\usepackage{indentfirst}
\usepackage{booktabs}

\begin{document}
	\renewcommand{\baselinestretch}{1.3}
	\newcommand\beq{\begin{equation}}
		\newcommand\eeq{\end{equation}}
	\newcommand\beqn{\begin{eqnarray}}
		\newcommand\eeqn{\end{eqnarray}}
	\newcommand\nn{\nonumber}
	\newcommand\fc{\frac}
	\newcommand\lt{\left}
	\newcommand\rt{\right}
	\newcommand\pt{\partial}

	\title{\Large{\bf Tidal Love numbers of Axion stars}}
	
	\author{ Jun-Ru Chen, Shi-Xian Sun, Long-Xing Huang, and  Yong-Qiang Wang\footnote{yqwang@lzu.edu.cn, corresponding author}
	}
	
	\affiliation{$^{1}$Lanzhou Center for Theoretical Physics, Key Laboratory of Theoretical Physics of Gansu Province,School of Physical Science and Technology,Lanzhou University, Lanzhou 730000, China,$^{2}$Institute of Theoretical Physics $\&$ Research Center of Gravitation,Lanzhou University, Lanzhou 730000, China}
	\date{\today}

	\begin{abstract}
		We investigate the tidal deformability of spherically symmetric axion stars on the stable branches, including the Newtonian and relativistic branches. The results suggest that on the stable branch, the electric Love numbers of axion star are positive, while the magnetic Love numbers are negative.
		On the Newtonian stable branch, the electric tidal Love numbers are much larger than the magnetic ones, while on the relativistic stable branch, they are slightly larger.
		Furthermore, the relativistic stable branch has much smaller tidal Love numbers than the Newtonian stable branch, indicating weaker deformability of axion stars on the relativistic stable branch.
		This could be attributed to the fact that on the relativistic branch, axion stars are more compact,  resulting hardly distorted by tidal forces.
	\end{abstract}

	\maketitle

	%%%%%%%%%%%%%%%%%%%%%%%%%%%%%%%%%%%%%%%%%%%%%%%%%%%%%%%%%%%%%%%%%%%%%%%%%%%%%%
	\section{Introduction}\label{Sec1}
	%%%%%%%%%%%%%%%%%%%%%%%%%%%%%%%%%%%%%%%%%%%%%%%%%%%%%%%%%%%%%%%%%%%%%%%%%%%%%%

	In an external tidal environment, the uneven gravitational fields and the relative motion between the bodies cause tidal effects, which are fundamental in astrophysics as they reveal the interactions and deformability of gravitational objects.
	Tidal effects can alter the rotational speeds of gravitational objects, giving rise to astrophysical phenomena such as tidal tails and tidal locking~\cite{goldreich1966spin,toomre1972galactic}.
	
	Ocean tide in the Earth-Moon system is the most well-known tidal phenomenon. It causes periodic rise and fall of the ocean surface, creates tidal friction and affects the Earth’s rotation rate, resulting in a slower rotation rate.
	To characterize Earth's response to tidal forces, A.~E.~H. Love introduced the concept of  Love numbers in Newtonian gravity, introducing two dimensionless parameters, denoted as $h$ and $k$.
	The $h$ describes the relative longitudinal deformation, while $k$ delineates the relative deformation in the gravitational potential.
	Later, the third dimensionless parameter, $l$, was introduced by T.~Shida to account for the relative horizontal deformation of the Earth, which is called Shida number sometimes~\cite{shida1912body}.
	Collectively, these three dimensionless parameters are recognized as the Love numbers~\cite{love1909yielding}.

	The prospects of observing the tidal Love numbers of spherically symmetric neutron stars have motivated the study of the tidal Love numbers in general relativity~\cite{Hinderer:2007mb,Flanagan:2007ix,Damour:2009vw,Binnington:2009bb}.
	Tidal deformation can affect the waveform of gravitational waves, which is related to the internal structure of neutron stars.
	By measuring the tidal Love numbers through gravitational wave observations, we can constrain the equation of state that governs the interiors of neutron stars~\cite{Flanagan:2007ix,Hinderer:2016eia,Dietrich:2020eud}.
	In the relativistic theory, the key deformation parameter is  $k$, which is related to the ratio between the multipole moments induced by deformations and the multipole moment of the tidal field.
	In terms of parity, Love numbers are categorized into even Love numbers (electric-type) and odd Love numbers (magnetic-type).
	The discovery of the I-Love-Q relations for slowly rotating neutron stars has provided a novel avenue for inferring the other two physical quantities through tidal Love numbers, thereby enabling tests of general relativity in the strong-field regime~\cite{Yagi:2013bca,Yagi:2013awa}.
	Furthermore, research efforts have extended to investigate tidal deformations in rotating compact objects~\cite{Pani:2015hfa,Pani:2015nua,Landry:2015snx,Landry:2015zfa,Landry:2017piv,Gupta:2020lnv,Landry:2018bil}.
	Additionally, calculations of tidal Love numbers have been explored within modified gravity  ~\cite{Yang:2022ees,Meng:2021ijp}.

	Afterward, the study about tidal deformation of black holes has yielded an intriguing result that for Schwarzschild black holes both their electric-type and magnetic-type Love numbers are found to be zero~\cite{Binnington:2009bb,Damour:2009vw,Fang:2005qq}.
	Subsequently, similar conclusions have been extended to slowly rotating black holes~\cite{Pani:2015hfa,Landry:2015zfa,Gurlebeck:2015xpa,
		Poisson:2014gka}.
	In other words, under the influence of tidal forces, there is no generation of induced multipole moments, thereby leaving the multi-polar structure of the black hole unaltered. Although the gravitational field of a black hole is extremely strong, due to its singularity and lack of actual material structure, it cannot generate tidal deformation, which is significantly different from other self-graviational objects.
	The tidal Love numbers within the gravitational wave signal during the binary inspiral can be used to test the properties of black holes and distinguish them from other compact objects to some extent~\cite{Cardoso:2017cfl,Sennett:2017etc,
		Cardoso:2019rvt}.

	Boson stars (BSs) as compact stars different from black holes are solitonic formed by self-gravitating bosonic fields.
	The theoretical framework for these configurations, involving Einstein gravity coupled to complex scalar field, was constructed by Kaup et al. and then R. Ruffini found stable solutions for such systems~\cite{Kaup:1968zz, Ruffini:1969qy}.
	In astrophysics, BSs are considered as candidates for dark matter~\cite{Hu:2000ke,Sahni:1999qe,Matos:2000ng,Hui:2016ltb}. The research findings indicate that the tidal Love numbers of BSs are smaller than those of neutron stars~\cite{Cardoso:2017cfl,
		Mendes:2016vdr}.
	Moreover, there have been studies computing tidal Love numbers for Proca Stars, which are vector bosonic stars, along with a rudimentary comparison with the results for scalar BSs. Spherical bosonic binary stars with the same compactness exhibit differences in the gravitational wave (GW) signals between scalar and vector during the inspiral~\cite{Herdeiro:2020kba}.
	Furthermore, investigations into the tidal Love numbers have extended to other exotic compact objects, including gravastars ~\cite{Uchikata:2016qku,
		Pani:2015tga} and quark stars ~\cite{Postnikov:2010yn,Albino:2021zml}.

	Axion stars as compact objects formed by axion fields,  consider a self-interaction complex scalar field minimally coupled to gravity, and are also considered one of the candidates for dark matter~\cite{Davidson:2016uok,Eby:2017xaw,Baer:2014eja,
		Klaer:2017ond}.
	In response to the strong CP problem in Quantum Chromo Dynamics (QCD), the Peccei-Quinn mechanism was proposed, introducing the axion as a novel particle ~\cite{Wilczek:1977pj,Peccei:1977hh,Weinberg:1977ma,Callan:1979bg}, considered as weakly-interacting ultralight bosons beyond Standard Model~\cite{Essig:2013lka}. Axion-like particles as particles beyond the Standard Model play a pivotal role in string theory models~\cite{Arvanitaki:2009fg,Svrcek:2006yi}.
	Spherically symmetric axion stars solutions and stability analysis were studied in~\cite{Guerra:2019srj} which found new stability branches emerging at high density.
	Recent studies have explored rotating axion stars~\cite{Delgado:2020udb} and the multi-field involving rotating axion stars mixed with boson fields~\cite{Zeng:2021oez}. Our work primarily focuses on the tidal deformability of spherically symmetric axion stars. Through numerical calculations, we have investigated the quadrupole and octupole tidal Love numbers of axion stars under various parameters $f_a$.

	The paper is organized as follows. In Sec.~\ref{sec2}, we briefly review the model of axion stars, considering complex scalar field minimally coupled to Einstein’s gravity. In Sec.~\ref{sec3}, we present the perturbation equations for axion stars, divided into odd and even perturbations. In Sec.~\ref{sec4}, we show the numerical results of quadrupole and octupole tidal Love numbers for axion stars under different decay constants. Finally, Sec.~\ref{sec5} is our conclusion and discussion. We adopt the signature $(-,+,+,+)$ for the metric and
	natural units $c=G=1$.

	\section{Spherically symmetric axion star}\label{sec2}
	
	\subsection{Model}
	
	Under the Einstein-Klein-Gordon (EKG) theory,  we consider the complex scalar field minimally couples with gravity, and the action is
	\begin{equation}
		\mathcal{S} = \int d^4 x \sqrt{-g} \left[ \frac{R}{16\pi} - g^{\alpha\beta} \partial_{\alpha} \Psi^* \partial_{\beta} \Psi - V(|\Psi|^2) \right] \ ,
		\label{Action}
	\end{equation}
	where $R$ represents the Ricci scalar, $\Psi$ denotes the axion field, and $V(|\Psi|^2)$ is the axion potential. From Eq.~\eqref{Action}, we derive the corresponding equations of motion. Taking the variation with respect to the metric $g_{\alpha\beta}$ yields the Einstein field equations
	\begin{equation}
		R_{\alpha\beta} - \frac{1}{2} g_{\alpha\beta} R = 8\pi \ T_{\alpha\beta}  \ .	
		\label{EinsteinEquation}
	\end{equation}
	Here, the energy-momentum tensor $T_{\alpha\beta}$ for the axion field is given by
	\begin{equation}
		T_{\alpha\beta} = \partial_{\alpha} \Psi^* \partial_{\beta} \Psi +\partial_{\beta} \Psi^* \partial_{\alpha} \Psi - g_{\alpha\beta} \left(g^{\mu\nu} \partial_\mu \Psi^* \partial_\nu \Psi + V\right) \ .
		\label{EnergyMomentumTensor}
	\end{equation}
	Variation with respect to the axion field $\Psi$ yields the Klein-Gordon equation
	\begin{equation}
		\Box \Psi  = \dfrac{\partial V}{\partial |\Psi|^2} \Psi \ . \label{KleinGordon}
	\end{equation}
	According to Noether's theorem, the action of a field remains invariant under U(1) transformations, where $\Psi \to \Psi e^{i\alpha}$, with $\alpha$ being a constant. This implies the existence of a conserved current
	\begin{equation}
		j^{\alpha} = -i \left( \Psi^* \partial^{\alpha} \Psi - \Psi \partial^{\alpha} \Psi^* \right) \ ,
		\label{Currents}
	\end{equation}
	which satisfies $\nabla_\alpha j^\alpha = 0$.		
	The integral of the timelike component of the 4-current on a spacelike hypersurface $\Omega$ yields a conserved quantity, the Noether charge
	\begin{equation}
		Q = \int_\Omega j^t \ .
		\label{Charge}
	\end{equation}
	We consider a spherically symmetric static background, within the following metric ansatz
	\begin{equation}
		ds^2 = -e^{\eta(r)} dt^2 + e^{\xi(r)} dr^2 + r^2 \left(d \theta^2+\sin^2 \theta d \varphi^2 \right)\ .
		\label{metrican}
	\end{equation}
	The metric is static, and the functions $\eta(r)$ and $\xi(r)$ depend only on the radial coordinate $r$. Using the ansatz of scalar field
	\begin{equation}
		\Psi^{(0)}=\psi_0(r) e^{-i \omega t} \ ,
		\label{ScalarField}
	\end{equation}
	where $\psi_0$ is a real scalar and $\omega$  represent the angular frequency of the scalar field.
	And the axion potential is
	\begin{equation}
		V(\psi_0) = \frac{2 \mu^2 f_a^2}{B} \left[ 1 - \sqrt{1 - 4 B \sin^2 \left( \frac{\psi_0}{2 f_a} \right)} \right] \ .
		\label{Potential}
	\end{equation}
	Here, $B$ is a constant related to the ratio of the up quark mass $m_u$ to the down quark mass $m_d$, with $m_u/m_d \approx 0.48$, giving $B \approx 0.22$, $\mu$ and $f_a$ are two free parameters. In this potential, the second term corresponds to the QCD axion effective potential~\cite{GrillidiCortona:2015jxo}, and the addition of a constant term ensures $V(0)=0$, in order to construct asymptotically flat axion stars. Expanding the potential around $ \psi_0=0 $
	\begin{equation}
		V(\psi_0) = \mu^2 \psi_0^2 - \left( \frac{3B-1}{12} \right) \frac{ \mu^2}{f_a^2} \psi_0^4 + \dots \ .
		\label{UnfoldingPotential}
	\end{equation}
	It can be observed that $\mu$ represents the mass of the axion, while $f_a$ denotes the decay constant of the axion field. When $f_a \gg \psi_0$, only the free scalar potential remains, and the axion star model reduces to the mini-boson stars~\cite{Ruffini:1969qy,Kaup:1968zz}.
	Taking the axion potential into Eq.~\eqref{EinsteinEquation} and Eq.~\eqref{KleinGordon}, we obtain a set of ordinary differential equations
	\begin{equation}
		\begin{gathered}
			\eta^{\prime}(r) =\frac{-1+e^{\xi(r)}}{r}+
			\frac{16 e^{\xi(r)} f_a^2 \pi r \mu^2\left(-1+\sqrt{1-2 B+2 B\cos \left(\frac{\psi_0(r)}{f_a}\right)}\right)}{B}+\\
			8 e^{-\eta(r)+\xi(r)} \pi r \omega^2 \psi_0(r)^2+8 \pi r \psi_0^{\prime}(r)^2 \ ,
			\label{differential1}
		\end{gathered}
	\end{equation}
	\begin{equation}
		\begin{gathered}
			\xi^{\prime}(r) =\frac{1-e^{\xi(r)}}{r}-
			\frac{16 e^{\xi(r)} f_a^2 \pi r \mu^2\left(-1+\sqrt{1-2 B+2 B\cos\left(\frac{\psi_0(r)}{f_a}\right)}\right)}{B}+ \\
			8 e^{-\eta(r)+\xi(r)} \pi r \omega^2 \psi_0(r)^2+8 \pi r \psi_0^{\prime}(r)^2 \ ,
			\label{differential2}
		\end{gathered}
	\end{equation}
	\begin{equation}
		\begin{gathered}
			\psi_0^{\prime \prime}(r)=\frac{e^{\xi(r)}f_a \mu^2 \sin \left(\frac{\psi_0(r)}{f_a}\right)}{\sqrt{1-2 B+2 B\cos\left(\frac{\psi_0(r)}{f_a}\right)}}-e^{-\eta(r)+\xi(r} \omega^2 \psi_0(r) - \\
			\frac{2 \psi_0^{\prime}(r)}{r}-\frac{1}{2} \eta^{\prime}(r) \psi_0^{\prime}(r)+\frac{1}{2} \xi^{\prime}(r) \psi_0^{\prime}(r) \ .
			\label{differential3}
		\end{gathered}
	\end{equation}
	These three equations does not have analytical expressions, but after applying appropriate boundary conditions, numerical solutions can be obtained.
	%%%%%%%%%%%%%%%%%%%%%%%%%%%%%%%%
	\subsection{Numerical solution}
	%%%%%%%%%%%%%%%%%%%%%%%%%%%%%%%%
	By providing appropriate boundary conditions, we can numerically solve Eq.~\eqref{differential1}-Eq.~\eqref{differential3}. At the origin,
	\begin{equation}
		\begin{gathered}
			\xi(0)=0 \ ,\qquad \eta(0)=\eta_c \ ,
			\\ \psi_0^{\prime}(0)=0\ ,\qquad \psi_0(0)=\psi_c \ ,
			\label{Origin}
		\end{gathered}
	\end{equation}
	where $\eta_c$ is arbitrary. At infinity, the scalar field vanishes,
	\begin{equation}
		\lim_{r\to \infty} \eta(r) = 0 \ ,\qquad  \lim_{r\to \infty} \psi_0(r) = 0 \ .
		\label{Infinity}
	\end{equation}
	In spherically symmetric static spacetime, when $ r \to\infty $ the mass of the axion star $M=m(r\to\infty)$ can be determined using the formula
	\begin{equation}
		m(r)=\frac{r}{2}(1-\frac{e^{-\eta(r)}}{r}) \ .
		\label{Mass}
	\end{equation}
	As axion stars do not have a rigid surface like neutron stars due to the scalar field  does not vanish at infinity, we define the effective radius R of an axion star as the radius that contains 99\% of the total mass~\cite{Schunck:2003kk}, i.e., $m(R)=0.99M$, and we define the compactness parameter of the axion star as $c=M/R$.
	
	It is convenient perform a radial coordinate transformation
	\begin{equation}
		x=\frac{R}{1+R} \ ,
		\label{}
	\end{equation}
	where the radial coordinate $R \in[0,\infty)$ and the new radial coordinate $x \in[0,1)$. Numerical solutions are obtained through the finite element method. The number of grid points are 1000 in the integration region $0 \leq x < 1$.
	To ensure the accuracy of the computational results, we impose a requirement that the relative error is less than $10^{-5}$.
	\begin{figure}[htbp]
		\centering
		\begin{minipage}[t]{0.96\textwidth}
			\includegraphics[width=\textwidth]{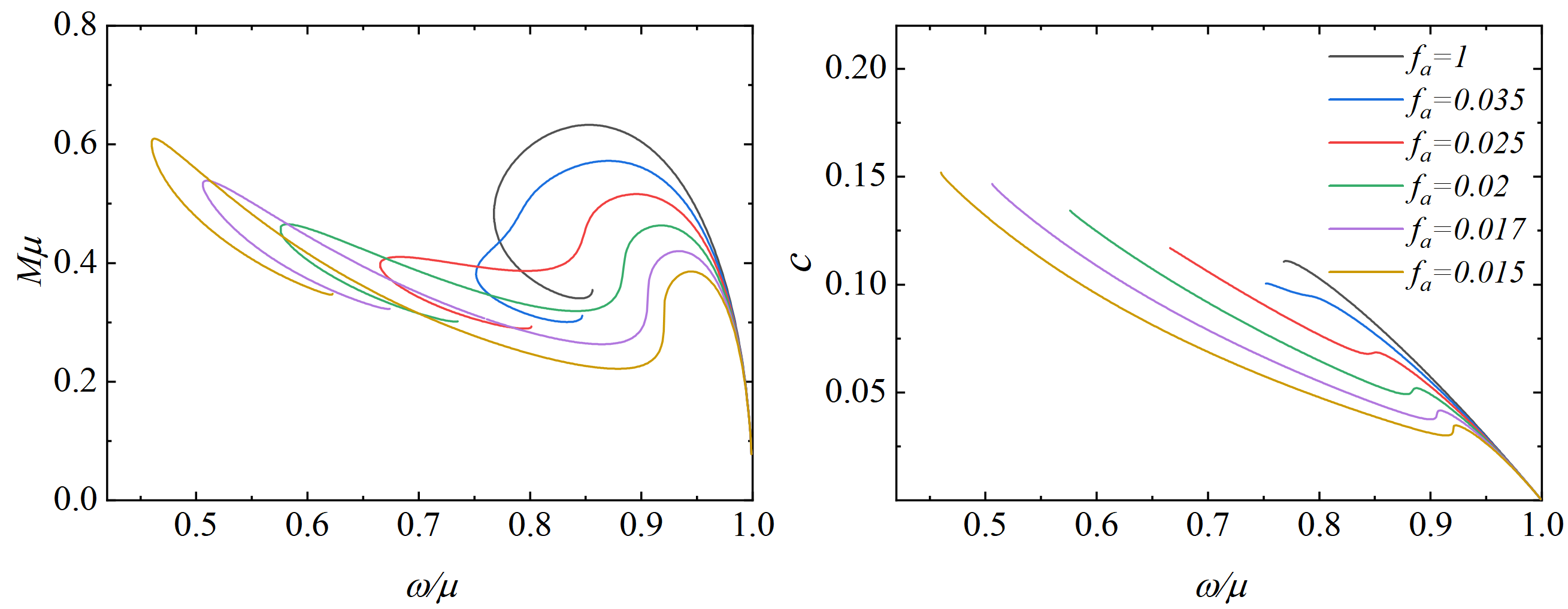}
		\end{minipage}
		\caption{Domain of existence of the axion star solutions for different decay constant $f_a$ in an ADM mass/compactness vs. frequency. Left panel: ADM mass $M$ vs. scalar field frequency $\omega$ diagram. Right panel: Compactness $c$ of axion star as a function of scalar field frequency $\omega$. Using solid lines of the same color represents the same value of $f_a$.}
		\label{fig:m-w}
	\end{figure}
	\begin{figure}[htbp]
		\centering
		\begin{minipage}[t]{0.8\textwidth}
			\includegraphics[width=\textwidth]{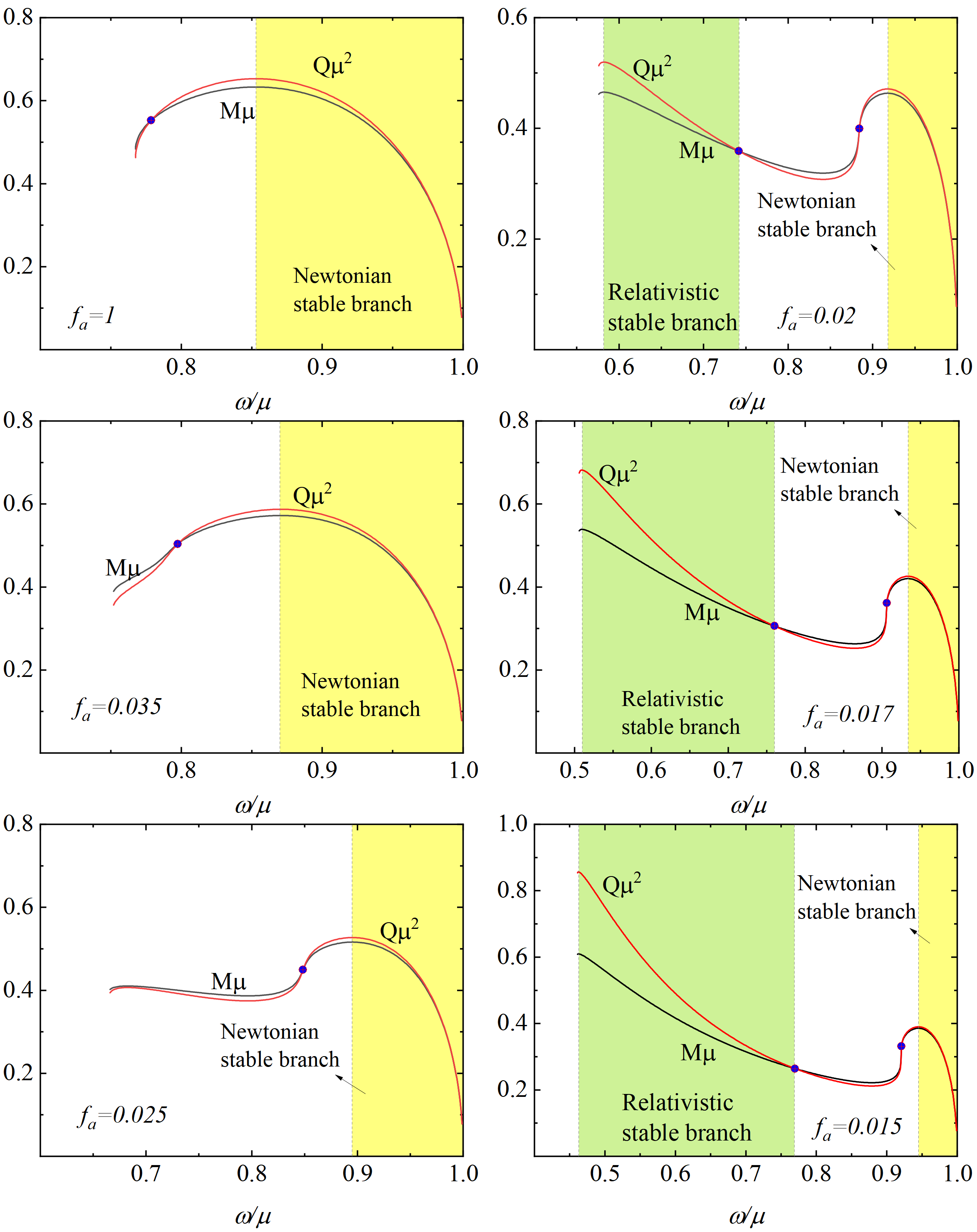}
		\end{minipage}
		\caption{Domain of existence of the axion star for both the ADM mass and Noether charge with $f_a=\{1,0.35,0.25,0.2,0.17,0.15\}$.The red solid line represents the Noether charge, while the black solid line represents the ADM mass.}
		\label{fig:m-q}
	\end{figure}
	
	The domain of existence of the axion star solutions for different decay constant $f_a$ in an ADM mass/compactness vs. frequency diagram is shown in Fig.~\ref{fig:m-w}.  Now, and for the remainder of this paper, we focus on axion stars with decay constant  $f_a=\{1,0.35,0.25,0.2,0.17,0.15\}$.
	The solid lines of different colors represent solutions for different $f_a$, illustrating the distribution of the total mass as a function of frequency in left panel.
	When $f_a=1$, the trend of the total mass with frequency is similar to mini-boson stars, suggesting that at larger values of $f_a$, axion stars decay to the mini-boson stars. For larger $f_a$, the variation of the total mass with frequency exhibits a spiral trend. As $f_a$ decreases, this spiral curve transforms into a shape like duck-bill.
	When $f_a$ is relatively small, the mass have two local maximum, suggesting the potential existence of two stable branches for axion stars.
	The right panel of Fig~.\ref{fig:m-w} shows the compactness of axion stars varies along the frequency for different $f_a$. The axion stars become more compact as the frequency decreases for larger $f_a$. However, when $f_a$ is small, the compactness initially increases with decreasing frequency, then starts decreasing before eventually increasing again.

	In order to characterize the region of the stable branch of axion star under these six chosen parameters $f_a$, we give domain of
	existence in
	Fig~.\ref{fig:m-q}, showing both the ADM mass and
	the Noether charge vs. the scalar field frequency.
	The red solid line represents the Noether charge, while the black solid line represents the total mass.
	We will study the tidal Love numbers under stable configurations.
	For mini-boson stars, the stable branch are determined from the maximum frequency, where the total mass $M$ tends to zero, to the maximum total mass $M$ ~\cite{Lee:1988av,Gleiser:1988rq}.
	However, the domain of stable solutions for axion stars differs from mini-boson stars.
	From the perspective of catastrophe theory, akin to axion stars with smaller $f_a$, exhibit two stable branches separated by an unstable region~\cite{Kleihaus:2011sx}.
	Axion stars with $f_a=0.02$ were numerically evolved in the study by Herdeiro et al.~\cite{Herdeiro:2021lwl}, confirming the existence of two stable branches.
	According to their result, when mass decreases with frequency and $Q \mu^2>M \mu$, the region is considered stable.
	For cases with $Q \mu^2<M \mu$, the solutions with energy  excess is unstable.
	The left panels of Fig.~\ref{fig:m-q} represents cases with one stable branch, highlighted in yellow.
	The right panels of Fig.~\ref{fig:m-q} represents cases with two stable branches. The stable region at higher frequencies corresponds to the Newtonian stable branch, highlighted in yellow, while the stable region at lower frequencies corresponds to the relativistic stable branch, highlighted in green. Fig.~\ref{fig:m-q} reveals that for larger $f_a$, only one stable branch exists.
	For smaller $f_a$ values, will exhibit two stable branches.
	Unlike mini-boson stars, axion stars have more compact and stable configurations that depend on different decay constant $f_a$.
	Axion stars have two stable branches when $f_a\leq0.024$, which is the critical parameter for the relativistic branch.

	\begin{figure}[htbp]
		\centering
		\includegraphics[width=0.96\linewidth]{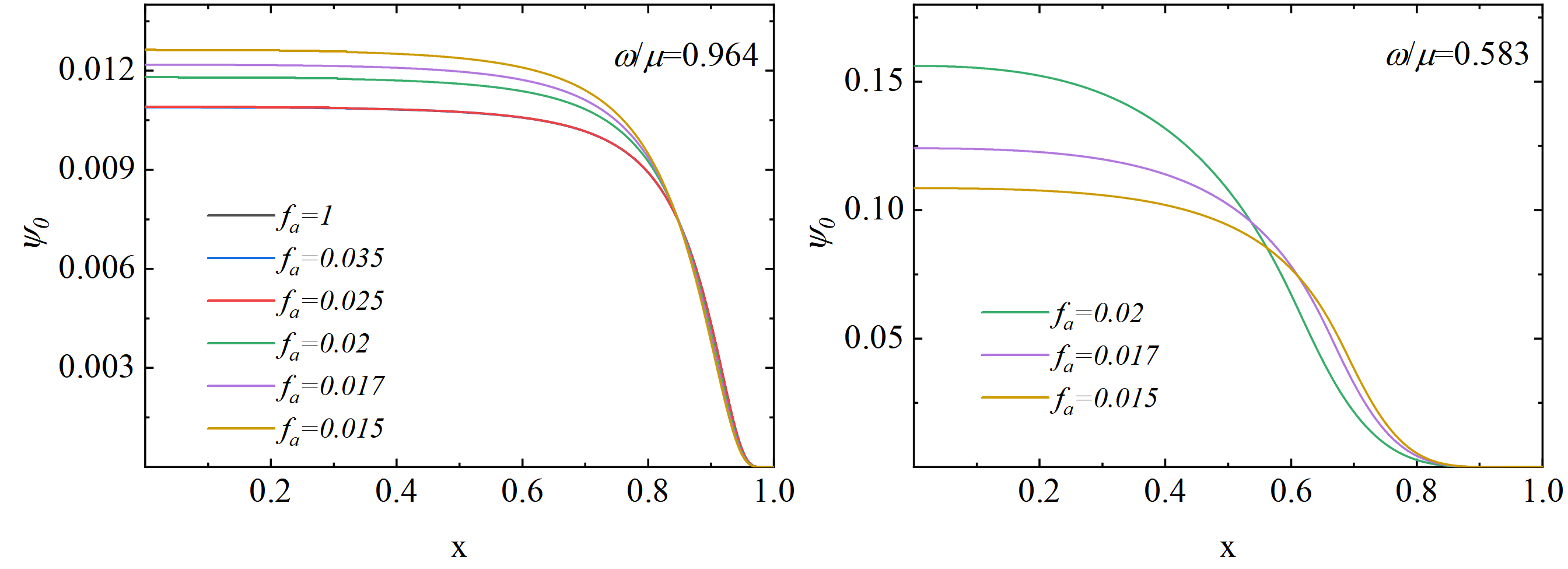}
		\caption{The distribution of the field function $\psi_0$ as a function of $x$ at $\omega/\mu=0.964$  (left panel)  and $\omega/\mu=0.583$ (right panel) for different $f_a$.
		}
		\label{fig:xf}
	\end{figure}
	
	The distribution of the scalar field $\psi_0$ as a function of radial coordinate $x$ are shown in Fig.~\ref{fig:xf}. The left panel illustrates the field function distribution on the Newtonian stable branch at $\omega/\mu=0.964$, while the right panel displays the field function distribution on the relativistic stable branch at $\omega/\mu=0.583$. Despite the distribution of field become very dilute when close infinity, it does not completely vanish.

	%%%%%%%%%%%%%%%%%%%%%%%%%%
	\section{Tidal perturbation}\label{sec3}
	%%%%%%%%%%%%%%%%%%%%%%%%%%
	
	We consider linear perturbations for axion stars immersed in a tidal field, specifically focusing on first-order linear perturbations. In a spherically symmetric background, the perturbed metric can be expressed as
	\begin{equation}
		g_{\alpha\beta}  = g_{\alpha\beta}^{(0)} + h_{\alpha\beta} \ ,
		\label{PerturbMetric}
	\end{equation}
	where $g_{\alpha\beta}^{(0)}$ is given by Eq.~\eqref{metrican}. The scalar field under perturbation is
	\begin{equation}
		\Psi = \Psi^{(0)} + \delta \Psi,
		\label{PerturbScalarField}
	\end{equation}
	with $\Psi^{(0)}$ defined in Eq.~\eqref{ScalarField}. The perturbation part of the scalar field is
	\begin{equation}
		\delta\Psi(t,r,\theta,\varphi)= \sum_{l,m} e^{-i\omega t}\psi_1(r)Y^{\ell m}(\theta,\varphi)\ .
		\label{PsiPerturb}
	\end{equation}
	The metric perturbations $h_{\alpha\beta}$ can fall into even-parity $h_{\alpha\beta}^{(e)}$ and odd-parity $h_{\alpha\beta}^{(o)}$
	\begin{equation}
		h_{\alpha\beta}  =  h_{\alpha\beta}^{(e)} + h_{\alpha\beta}^{(o)}  \ .
		\label{MetricPerturbations}
	\end{equation}
	In the Regge-Wheeler gauge, $h_{\alpha\beta}^{(e)}$ and $h_{\alpha\beta}^{(o)}$ can be expressed as\cite{Regge:1957td}
	\begin{equation}
		h_{\alpha\beta}^{\text (e)}=\left(\begin{array}{cccc}
			e^{\eta(r)} H_0^{\ell m}(r)  & 0  & 0 & 0
			\\
			0  & e^{\xi(r)} H_2^{\ell m}(r)  & 0 & 0
			\\
			0 & 0 & r^2 K^{\ell m}(r)  & 0
			\\
			0 & 0 & 0 & r^2 \sin ^2 \theta K^{\ell m}(r)
		\end{array}\right)Y^{\ell m} (\theta,\varphi),
		\label{Heven}
	\end{equation}
	
	\begin{equation}
		h_{\alpha\beta}^{\text (o)}=\left(\begin{array}{cccc}
			0 & 0 & h_0^{\ell m}(r) F_\theta^{\ell m} & h_0^{\ell m}(r) F_\varphi^{\ell m}
			\\
			0 & 0 & h_1^{\ell m}(r) F_\theta^{\ell m} & h_1^{\ell m}(r) F_\varphi^{\ell m}
			\\
			h_0^{\ell m}(r) F_\theta^{\ell m} & h_1^{\ell m}(r) F_\theta^{\ell m} & 0 & 0
			\\
			h_0^{\ell m}(r) F_\varphi^{\ell m}  & h_1^{\ell m}(r) F_\varphi^{\ell m}  & 0 & 0
		\end{array}\right),
		\label{Hodd}
	\end{equation}
	Where $F_\theta^{\ell m} = -\frac{1}{\sin\theta}\frac{\partial Y^{\ell m}(\theta,\varphi)}{\partial \varphi}$ and $F_\varphi^{\ell m} = \sin\theta \frac{\partial Y^{\ell m}(\theta,\varphi)}{\partial \theta}$. Here, $Y^{\ell m}$ represents spherical harmonics, and for simplicity, we consider $m=0$ in our subsequent calculations.
	The Einstein field equations and Klein-Gordon equation under linear perturbations are given by~\cite{Cardoso:2019rvt}
	\begin{equation}
		\delta G_{\beta}^{\alpha}  = 8 \pi \delta T_{\beta}^{\alpha} \ ,
		\label{EinsteinPerturb}
	\end{equation}
	\begin{equation}
		\delta( \nabla^\alpha \nabla_\alpha \Psi-\dfrac{\partial V}{\partial |\Psi|^2} \Psi) = 0 \ .
		\label{KGPerturb}
	\end{equation}
	The main objective of this paper is to describe the tidal deformability of axion stars in a tidal field by solving the perturbed Einstein field equations. For non-rotating axion stars, neglecting odd-even perturbation mixing, we can classify tidal deformations into electric-type and magnetic-type. Electric-type tidal deformations have even parity, while magnetic-type tidal deformations have odd parity. The induced multipole moments are related to the deformation and mass distribution of axion stars, and the tidal Love numbers can be determined through the relationship between multipole moments and induced multipole moments. Considering linear perturbations, the induced multipole moments generated inside axion stars due to tidal deformations are related to the external tidal field's multipole moments. The relationship between $\ell$ order tidal multipole moments is
	\begin{equation}
		\begin{aligned}
			M_\ell & =\lambda_{\ell} \tilde{E}_\ell \ , \\
			S_\ell & =\sigma_{\ell} \tilde{B}_\ell \ .
		\end{aligned}
	\end{equation}
	Where $\lambda_{\ell}$ and $\sigma_{\ell}$ are the tidal-polarizability coefficients~\cite{Damour:1991yw}. ~$M_\ell$ and $S_\ell$ represent the mass and spin multipole moments generated, respectively, and $\tilde{E}_\ell$ and $\tilde{B}_\ell$ represent the electric and magnetic multipole moments of the external tidal field.
	
	To further calculate the tidal Love numbers, we first need to obtain the multipole moments of the tidal field and the induced multipole moments caused by the tidal field. These two types of multipole moments can be extracted through an asymptotic behavior of the metric. We adopt the multipole moment extraction method proposed by Kip S. Thorne~\cite{Thorne:1980ru}. In this method, an asymptotically Cartesian mass-centered coordinate system (ACMC) is chosen, and the $(t,t)$ component and $(t,\varphi)$ component of the metric can be expressed as
	\begin{equation}
		\begin{aligned}
			& g_{t t}=-1+\frac{2 M}{r}+\sum_{\ell \geq 2}\left(\frac{2}{r^{\ell+1}}\left[\sqrt{\frac{4 \pi}{2 \ell+1}} M_\ell Y^{\ell 0}+\left(\ell^{\prime}<\ell \text { pole }\right)\right]-\frac{2}{\ell(\ell-1)} r^l\left[\tilde{E}_\ell Y^{\ell 0}+\left(\ell^{\prime}<\ell \text { pole }\right)\right]\right) \ , \\
			& g_{t \varphi}=\frac{2 J}{r} \sin ^2 \theta+\sum_{\ell \geq 2}\left(\frac{2}{r^\ell}\left[\sqrt{\frac{4 \pi}{2 \ell+1}} \frac{S_\ell}{\ell} F_\varphi^{\ell 0}+\left(\ell^{\prime}<\ell \text { pole }\right)\right]+\frac{2 r^{\ell+1}}{3 \ell(\ell-1)}\left[\tilde{B}_l F_\varphi^{\ell 0}+\left(\ell^{\prime}<\ell \text { pole }\right)\right]\right) \ .
		\end{aligned}
		\label{GExpand}
	\end{equation}
	
	Tidal Love numbers are related to the asymptotic behavior of the metric. From the tidal deformability coefficients $ \lambda_{\ell}$ and $ \sigma_{\ell} $, we define dimensionless electric-type tidal Love numbers as $k^E_{\ell}$ and magnetic-type tidal Love numbers as $k^B_{\ell}$~\cite{Cardoso:2017cfl}
	\begin{equation}
		\begin{split}
			&k^E_{\ell}\equiv -\frac12 \frac{\ell(\ell-1)}{M^{2\ell+1}} \sqrt{\frac{4\pi}{2\ell+1}}\frac{M_\ell}{\tilde{E}_\ell} \ ,\\
			& k^B_{\ell}\equiv -\frac32 \frac{\ell(\ell-1)}{(\ell+1)M^{2\ell+1}}\sqrt{\frac{4\pi}{2\ell+1}}  \frac{S_\ell}{\tilde{B}_\ell} \ .
		\end{split}
		\label{TLN}
	\end{equation}
	Here, $M$ represents the mass of the axion star, and we use  $M^{2\ell+1}$ to make the tidal Love numbers dimensionless.

	%%%%%%%%%%%%%%%%%%%%%%%%%%%
	\subsection{Electrical perturbation}
	%%%%%%%%%%%%%%%%%%%%%%%%%%%
	
	We first consider the electric-type tidal perturbations produced by the external tidal field. By substituting the scalar field perturbation from Eq.~\eqref{PsiPerturb} and the even perturbation from Eq.~\eqref{Heven} into the linearized Einstein equation from Eq.~\eqref{EinsteinPerturb}, combining the $(\theta,\theta)$ and $(\varphi,\varphi)$ components of the linearized Einstein equation, we obtain $H_2(r) = H_0(r)$. Using the $(r,\theta)$ component, we can express $K'(r)$ as a function of $H_0(r)$ and $\psi_1(r)$
	\begin{equation}
		K'(r) = H_0'(r) + H_0(r) \eta'(r) - 32\pi \psi_1(r) \psi'_0(r) \ .
		\label{DerivativeK}
	\end{equation}
	Here, $\psi_0(r)$ is obtained from the background solution. By using $K'(r)$ and $H_2(r)=H_0(r)$, we can derive a linear equation for $H_0(r)$ by subtracting the $(t,t)$ component from the $(r,r)$ component of the linearized Einstein equation
	\begin{equation}
		\begin{gathered}
			a_1 H_0 +a_2 H_0^{\prime}+	H_0^{\prime\prime}
			=a_3 \psi_1 \ .
			\label{FunctionH}
		\end{gathered}
	\end{equation}
	Here, $a_1$, $a_2$, and $a_3$ depend on the background solution and are given by
	\begin{equation}
		\begin{gathered}
			a_1=32 e^{-\eta+\xi} \pi \omega^2 \psi_0^2-\frac{\left(\ell^2+\ell\right) e^{\xi}}{r^2}+ \frac{ 2 \eta^{\prime}}{r}-
			\frac{\eta^{\prime 2}}{2} -
			\frac{\eta^{\prime} \xi^{\prime}}{2} +\eta^{\prime \prime} \ ,
			\\
			a_2=\frac{2}{r}+\frac{\eta^{\prime}-\xi^{\prime}}{2} \ ,
			\\
			a_3 = 32 \pi \left[ -e^{-\eta+\xi}  \omega^2 \psi_0+\psi_0^{\prime}\left( \frac{2}{r}-\frac{\eta^{\prime}+\xi^{\prime}}{2 } \right)+ \psi_0^{\prime \prime}\right] \ .
			\label{CoefficientA}
		\end{gathered}
	\end{equation}
	Obtaining the equation for $\psi_1(r)$ from the Klein Gordon equation under perturbation
	\begin{equation}
		\begin{gathered}
			b_1 \psi_1+ b_2 \psi_1^{\prime}+ \psi_1^{\prime \prime}
			= b_3 H_0 \ .
			\label{FunctionPsi}
		\end{gathered}
	\end{equation}
	Here, $b_1$, $b_2$, and $b_3$ depend on the background solution and are given by
	\begin{equation}
		\begin{gathered}
			b_1=-\frac{\left(\ell^2+\ell\right) e^{\xi}}{r^2}+e^{-\eta+\xi} \omega^2-\frac{e^{\xi} \mu^2 \operatorname{\cos}\left(\frac{\psi_0}{f_ a}\right)}{\sqrt{1-2 B+2B\operatorname{\cos}\left(\frac{\psi_0}{f_a}\right)}}-
			\frac{B e^{\xi} \mu^2 \operatorname{\sin}^2\left(\frac{\psi_0}{f_ a}\right)}{\left[1-2 B+2 B \operatorname{\cos}\left(\frac{\psi_0}{f_ a}\right)\right]^{3 / 2}}-32 \pi \psi_0^{\prime 2} \ ,
			\\
			b_2=\frac{2}{r}+\frac{\eta^{\prime}}{2}-\frac{\xi^{\prime}}{2} \ ,
			\\
			b_3=-e^{-\eta+\xi} \omega^2 \psi_0+\frac{2 \psi_0^{\prime}}{r}-\frac{\left(\eta^{\prime}+\xi^{\prime}\right) \psi_0^{\prime}}{2}+\psi_0^{\prime \prime} \ .
			\label{CoefficientB}
		\end{gathered}
	\end{equation}
	By imposing appropriate boundary conditions, we can solve the system of equations in Eq.~\eqref{FunctionH} and Eq.~\eqref{FunctionPsi}. To improve the numerical behavior of the perturbation equations near the boundaries and enhance the reliability and accuracy of the calculations, we applied transformations to $H_0$ and $\psi_1$.
	\begin{equation}
		\tilde{H}_0(r) \equiv H_0 r^{-\ell}\ ,\quad
		\tilde{\psi}_1(r) \equiv \psi_1 r^{-(\ell+1)} \ .
	\end{equation}
	At the origin $r=0$, the boundary conditions are given by
	\begin{equation}
		\begin{gathered}
			\tilde{H}_0(0)=\tilde H_0^{(\ell)}\ ,\quad \tilde{H}_0^{\prime}(0)=0 \ ,
			\\
			\tilde{\psi}_1(0) =\tilde\psi_1^{(\ell+1)} \ ,\quad \tilde{\psi}_0^{\prime}(0)=0 .
			\label{Origin}
		\end{gathered}
	\end{equation}
	Since the system is linear, we can choose a specific value for $\tilde H_0^{(\ell)}$, such as $\tilde H_0^{(\ell)}=1$. During the numerical solution, we adjust $\tilde\psi_1^{(\ell+1)}$ so that $\tilde\psi_1 \to 0$ as $r \to \infty$ to determine the value of $\tilde\psi_1^{(\ell+1)}$.
	When we perform calculations at a extraction radius $R_\text{ext}$ which is far from away the center of axion stars, Eq.~\eqref{FunctionH} reduces to
	\begin{equation}
		H_0^{\prime\prime} +\left(\frac{2}{r}+\frac{\eta^{\prime}-\xi^{\prime}}{2}\right) H_0^{\prime}+\left(-\frac{\left(\ell^2+\ell\right) e^{\xi}}{r^2}+ \frac{ 2 \eta^{\prime}}{r}- \frac{\eta^{\prime 2}}{2} -
		\frac{\eta^{\prime} \xi^{\prime}}{2} +\eta^{\prime \prime}\right) H_0
		=0 \ .
		\label{LargeH}
	\end{equation}
	Using $e^{\nu(r)}=1-2M/r=e^{-\lambda(r)}$, we obtain
	\begin{equation}
		H_0'' + \frac{2(r-M)}{r(r-2M)} H_0' +\frac{ r (\ell^2+\ell)  (2M-r)-4 M^2} {r^2 (r-2 M)^2} H_0 = 0\ ,
		\label{LargeH0}
	\end{equation}
	We introduce the independent variable $f=r/M-1$, and Eq.~\eqref{LargeH0} has a general solution
	\begin{equation}
		H_0 = E_p P_\ell^2(r/M-1) + E_q Q_\ell^2(r/M - 1) \ ,
	\end{equation}
	where $P_\ell^2$ represents the Legendre function and $Q_\ell^2$ represents the associated Legendre function of the second kind. $E_p$ and $E_q$ are two integration constants, which can be determined by comparing the behavior of $H_0$ with that of the metric as $r\to\infty$. As $r\to \infty$, the asymptotic forms are
	\begin{equation}
		P_\ell^2 \approx a_p(\ell,M) \left(\frac{r}{M}\right)^\ell \ ,\quad Q_\ell^2 \approx a_q(\ell,M) \left(\frac{M}{r}\right)^{\ell+1} \ .	
	\end{equation}
	Using Eq.~\eqref{TLN}, we can express the electrical tidal Love numbers as
	\begin{equation}
		k^E_{\ell}=\frac12\frac{1}{M^{2\ell+1}}\frac{E_q a_q(\ell,M)}{E_p a_p(\ell,M)}\ ,
	\end{equation}
	When calculating the tidal Love numbers at $R_\text{ext}$, we define a new function
	\begin{equation}
		y=R_\text{ext} \frac{H_0^{\prime}(R_\text{ext})}{H_0(R_\text{ext})} \ .
	\end{equation}
	For $\ell = 2,3$, the electric tidal Love numbers are given by

	\begin{equation}
		\begin{aligned}
			k_2 & =\frac{8}{5}(1-2 \mathcal{C})^2 [2 \mathcal{C}(y-1)-y+2] \\
			& \times\left\{2 \mathcal{C}\left(4(y+1) \mathcal{C}^4+(6 y-4) \mathcal{C}^3+(26-22 y) \mathcal{C}^2+3(5 y-8) \mathcal{C}-3 y+6\right)\right. \\
			& \left.-3(1-2 \mathcal{C})^2(2 \mathcal{C}(y-1)-y+2) \log \left(\frac{1}{1-2 \mathcal{C}}\right)\right\}^{-1}, \\
			k_3 & =\frac{8}{7}(1-2 \mathcal{C})^2\left[2(y-1) \mathcal{C}^2-3(y-2) \mathcal{C}+y-3\right] \\
			& \times\left\{2 \mathcal{C}\left[4(y+1) \mathcal{C}^5+2(9 y-2) \mathcal{C}^4-20(7 y-9) \mathcal{C}^3+5(37 y-72) \mathcal{C}^2-45(2 y-5) \mathcal{C}+15(y-3)\right]\right. \\
			& \left.-15(1-2 \mathcal{C})^2\left(2(y-1) \mathcal{C}^2-3(y-2) \mathcal{C}+y-3\right) \log \left(\frac{1}{1-2 \mathcal{C}}\right)\right\}^{-1} \cdot	
		\end{aligned}
	\end{equation}
	where $\mathcal{C}=M/R_\text{ext}$.  $k_2$ and $k_3$ are quadrupolar and octupolar electric tidal Love numbers, respectively.

	%%%%%%%%%%%%%%%%%%%%%%%%%%%
	\subsection{Magnetic perturbation}
	%%%%%%%%%%%%%%%%%%%%%%%%%%%
	
	Next, we consider the magnetic tidal perturbations in the tidal environment. Using Eq.~\eqref{PsiPerturb} and Eq.~\eqref{Hodd} in Eq.~\eqref{EinsteinPerturb}, we get $\psi_0=0$ from the $(r,\theta)$ component and the equation for $h_0$ from the $(t,\varphi)$ component
	\begin{equation}
		c_1 h_0(r)+c_2 h_0^{\prime}(r)+	h_0^{\prime \prime}(r)=0 \ ,
		\label{Functionh}
	\end{equation}
	where $c_1$ and $c_2$ are defined as
	\begin{equation}
		\begin{gathered}
			c_1=\frac{-e^{\xi}\left(l^2+l-2\right)+r \left(\eta^{\prime}+\xi^{\prime}\right)-2}{r^2} \ ,
			\\
			c_2=-\frac{\xi^{\prime}+\eta^{\prime}}{2} \ .
			\label{CoefficientC}
		\end{gathered}
	\end{equation}
	We perform a transformation on $h_0$ to remove its explicit dependence on $r$
	\begin{equation}
		\tilde{h}_0(r)\equiv h_0 r^{l+1} \ .
	\end{equation}
	Under this transformation, the boundary conditions at the origin $r=0$ become
	\begin{equation}
		\tilde{h}_0(0)=\tilde{h}_0^{(\ell+1)} \ , \quad \tilde{h}_0^{\prime}(0)=0 \ .
	\end{equation}
	When extracting at a radius $\text{ext}$ much larger than the effective radius $R$, and utilizing $\xi^{\prime}=\frac{2M}{r(2M-r)}=-\eta$, Eq.~\eqref{Functionh} simplifies to
	\begin{equation}
		\frac{4M-\ell (\ell+1)}{r^2(r-2M)} h_0^{\prime}(r)+	h_0^{\prime \prime}(r)=0 \ .
	\end{equation}
	The solution to this differential equation is
	\begin{equation}
		\begin{aligned}
			h_0 & =B_f \frac{r^2}{4M^2} {}_2F_1\left(1-\ell, 2+\ell ; 4 ; \frac{r}{2 M}\right)  +B_g G_{2\ 2}^{2\ 0}\left(\frac{r}{2 M}
			\begin{vmatrix} 1-\ell,&2+\ell\\-1,&2 \end{vmatrix} \right) \ .
		\end{aligned}
	\end{equation}
	Here, $B_f$ and $B_g$ are integration constants, $_2F_1$ represents the hypergeometric function, and $G_{2\ 2}^{2\ 0}$ represents the Meijer function. As $r\to \infty$, the asymptotic behavior is
	\begin{equation}
		h_0 \approx B_f b_f(\ell,M) \left(\frac{r}{M}\right)^{\ell+1} + B_g b_g(\ell,M) \left(\frac{M}{r}\right)^\ell \ .	
	\end{equation}
	Using Eq.~\eqref{TLN}, we can express the magnetic tidal Love numbers as
	\begin{equation}
		k^B_{\ell}=-\frac12\frac{\ell}{\ell+1}\frac{1}{M^{2\ell+1}}\frac{B_g b_g(\ell,M)}{B_f b_f(\ell,M)}\ .
	\end{equation}
	By combining this with the asymptotic behavior of the metric's $(t,\varphi)$ component, we can derive the expressions for $k_2^B$ (quadrupolar) and $k_3^B$ (octupolar) magentic tidal Love numbers when $\ell=2,3$ as follows
	\begin{equation}
		\begin{aligned}
			k_2^B & =\frac{8}{5} \frac{2 \mathcal{C}(y-2)-y+3}{2 \mathcal{C}\left[2 \mathcal{C}^3(y+1)+2 \mathcal{C}^2 y+3 \mathcal{C}(y-1)-3 y+9\right]+3[2 \mathcal{C}(y-2)-y+3] \log (1-2 \mathcal{C})} \ , \\
			k_3^B=\frac{8}{7} & \left(8 \mathcal{C}^2(y-2)-10 \mathcal{C}(y-3)+3(y-4)\right)\left(15\left[8 \mathcal{C}^2(y-2)-10 \mathcal{C}(y-3)+3(y-4)\right] \log (1-2 \mathcal{C})\right. \\
			& \left.+2 \mathcal{C}\left[4 \mathcal{C}^4(y+1)+10 \mathcal{C}^3 y+30 \mathcal{C}^2(y-1)-15 \mathcal{C}(7 y-18)+45(y-4)\right]\right)^{-1} \ .
		\end{aligned}
	\end{equation}
	where $y=R_\text{ext} h_0^{\prime}(R_\text{ext})/h_0(R_\text{ext})$ and $\mathcal{C}=M/R_\text{ext}$.
	
	%%%%%%%%%%%%%%%%%%%%%%%%%%%
	\section{Tidal love number of Axion Stars}\label{sec4}
	%%%%%%%%%%%%%%%%%%%%%%%%%%%

	In this section, we will study the tidal Love numbers of spherically symmetric axion stars by computing y and $\mathcal{C}$ at extraction radius $R_\text{ext}$.
	It is noted that in Section~\ref{sec2} the first three parameters, characterized by larger values of $f_a$, exhibit only one stable branch known as the Newtonian stable branch. Conversely, for the latter three cases with smaller $f_a$, the system displays the presence of two stable branches.
	The branch from the maximum frequency to the first local maximum mass corresponds to the Newtonian stable branch, while the relativistic stable branch corresponds to the other stable branch.

	\begin{figure}[htbp]
		\centering
		\includegraphics[width=0.96\textwidth]{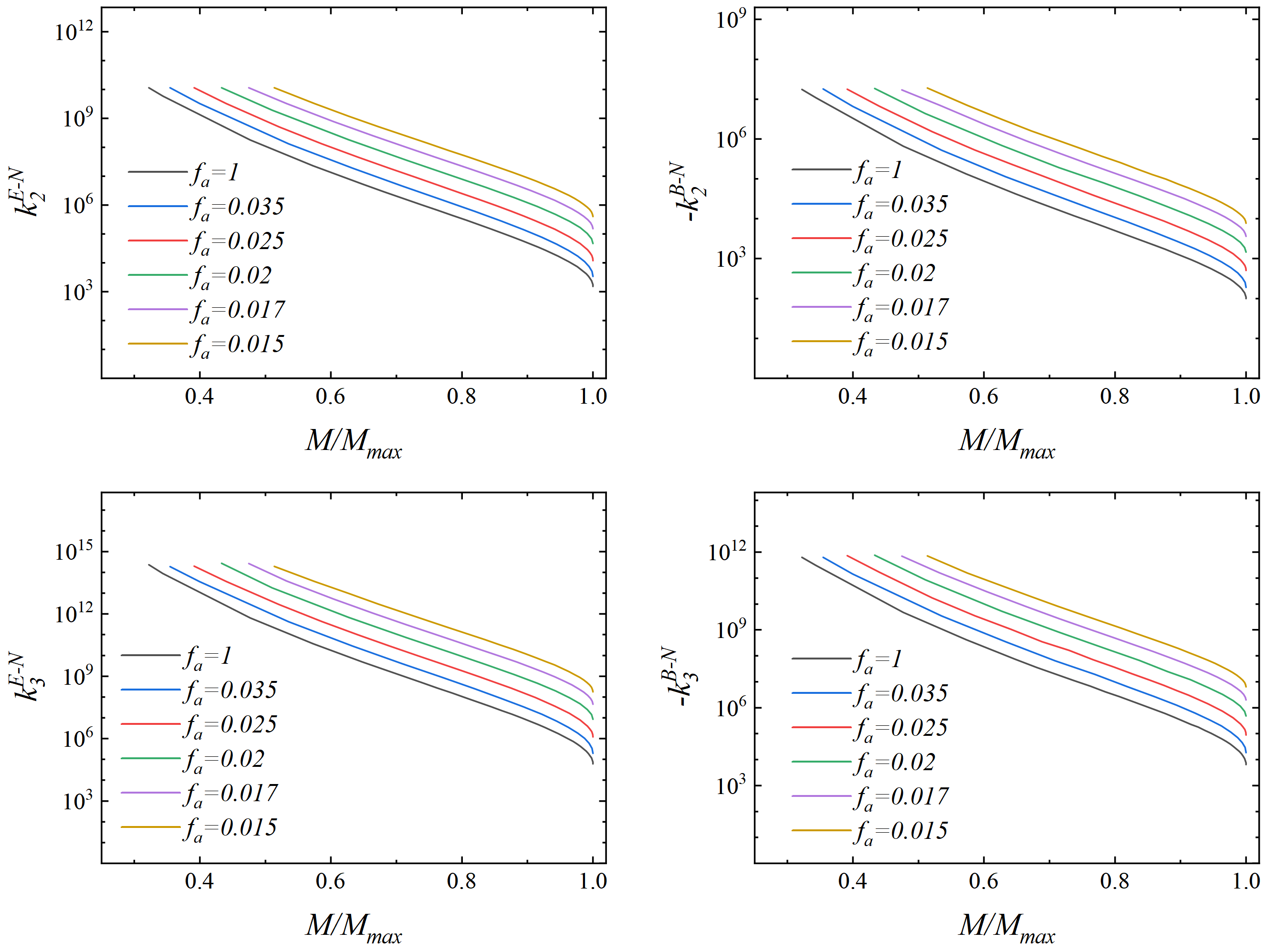}
		\caption{Tidal Love numbers for axion stars on the Newtonian stable branch, including the electric-type (left) and magnetic-type (right). The top panels represent the quadrupolar tidal Love numbers ($\ell=2$), while the bottom panels depict the octupolar tidal Love numbers ($\ell=3$).
		}
		\label{fig:love}
	\end{figure}

	\begin{figure}[htbp]
		\centering
		\includegraphics[width=0.96\textwidth]{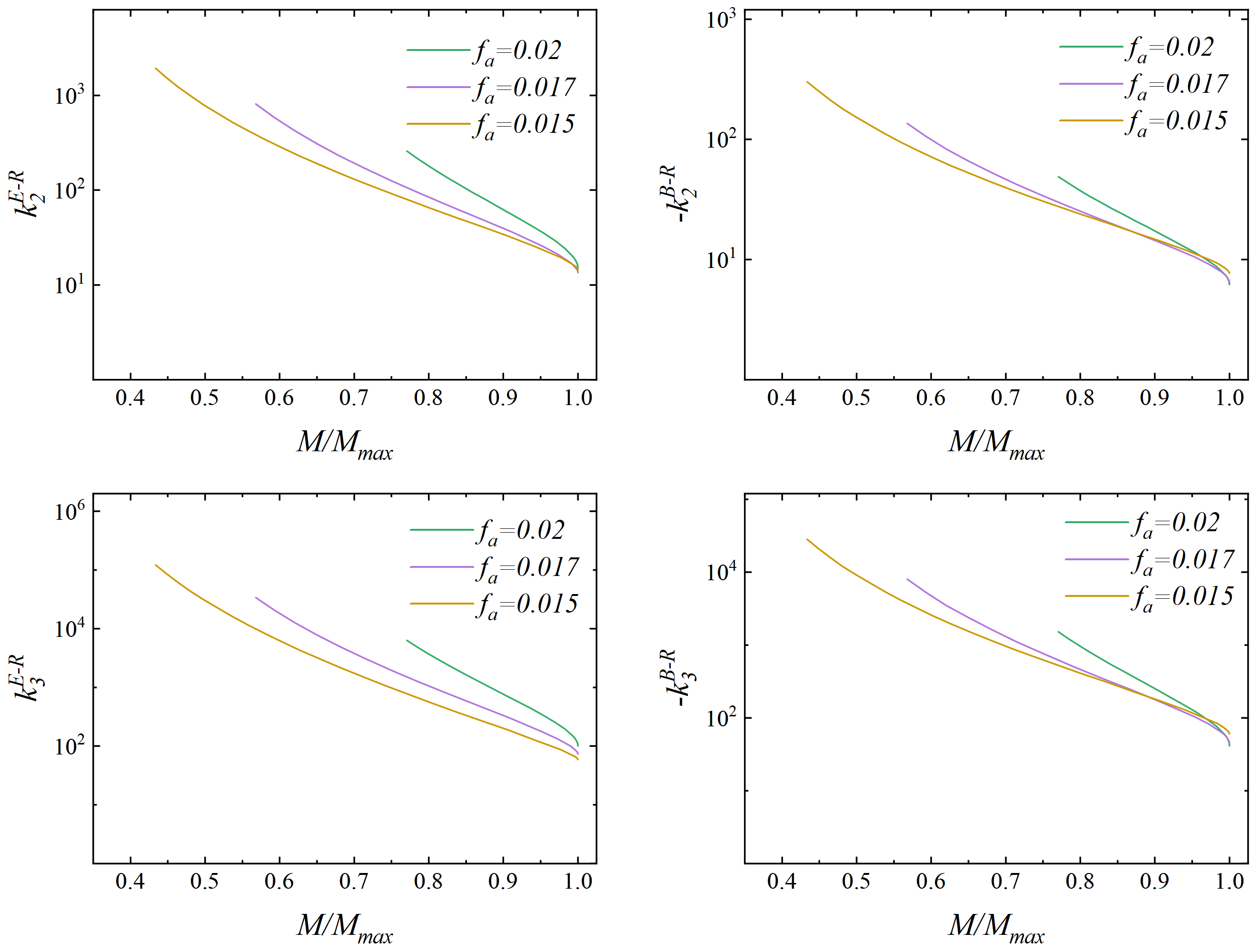}
		\caption{Tidal Love numbers for axion stars on the relativistic stable branch, including the electric-type (left) and magnetic-type (right). The top panels represent the quadrupolar tidal Love numbers ($\ell=2$), while the bottom panels depict the octupolar tidal Love numbers ($\ell=3$).}
		\label{fig:lovere}
	\end{figure}
	
	First, we analyze tidal Love numbers on the Newtonian stable branch. We calculate both the quadrupolar Love number ($\ell=2$) and the octupolar Love number ($\ell=3$). In Fig.~\ref{fig:love}, the left panels show the electric tidal Love number $k_\ell ^{E}$, while the right panels exhibit the magnetic tidal Love number $k_\ell ^{B}$ as functions of the axion star mass $M$ for different parameter $f_a$. The top panels correspond to the quadrupole case ($\ell=2$), whereas the bottom panels repersent to the octupole case ($\ell=3$).
	
	On the Newtonian stable branch for $\ell=2$ , we observe that the electric tidal Love number $k^{E-N}_{2}$ decreases with increasing mass under the same parameter $f_a$. As the mass increases, the axion stars become very compact, and compactness reach maximum at $M/M{\text{max}}=1$, where $k^{E-N}_{2}$ reaches its minimum. On the other hand, the absolute value of the magnetic tidal Love number $k^{B-N}_{2}$ decreases with increasing mass.
	The variation in the parameter $f_a$ does not alter the general trend of the tidal Love numbers with mass.
	
	For a constant mass ratio $M/M_{\text{max}}$, the self-interaction of the axion potential becomes stronger and quadrupole tidal Love numbers becomes larger  as the parameter $f_a$ decreases, suggesting increased tidal deformability as $f_a$ decreases.
	The electric tidal Love numbers are positive on this branch suggesting a positive feedback from the additional potential induced by the external gravitational field, promoting deformation.
	The negative magnetic tidal Love numbers show that the external tidal field produces a deformation which results in a spin multipole moment that is opposite to the multipole moment of the external tidal field.
	At a consistent mass ratio, the magnitude of the magnetic tidal Love numbers are observed to be smaller than that of the electric tidal Love numbers. We also perform calculations for the case of $\ell=3$ and obtain results similar to those for $\ell=2$.
	
	Next, we investigate the situation on the relativistic stable branch of axion stars. Tidal Love numbers for axion stars on the relativistic stable branch are presented in Fig.~\ref{fig:lovere}. The trends of tidal Love numbers with respect to mass on the relativistic stable branch closely resemble those observed on the Newtonian stable branch. The electric tidal Love number $k^{E-R}_{2}$, and the absolute value of the magnetic tidal Love number $k^{B-R}_{2}$ decrease as the mass increases. On this branch, the electric tidal Love numbers are positive, and the magnetic tidal Love numbers are negative. However, in contrast to the Newtonian stable branch, on the relativistic stable branch, for the same mass ratio, the magnitudes of the tidal Love numbers $k^{E-R}_{2}$ and $k^{B-R}_{2}$ are comparable, especially as $M/M_{\text{max}}\to1$, where they become very close.  Regardless of sign, for the same parameter and mass ratio, the electric tidal Love numbers are always greater than magnetic tidal Love numbers.  Furthermore, the Love numbers on the relativistic stable branch are markedly smaller than those on the Newtonian stable branch, indicating significantly weaker tidal deformability of axion stars on the relativistic stable branch compared to the Newtonian stable branch. At $M/M_{\text{max}}=1$, $k^{E-R}_{2}$ and $k^{B-R}_{2}$ tend to approach zero. This may be attributed to the more compact configurations of axion stars on the relativistic stable branch, resulting hardly distorted compared to those on the Newtonian stable branch.

	%%%%%%%%%%%%%%%%%%%%%%%%%%%%%%%%%%%%%%%%%%%
	\section{Conclusions}\label{sec5}
	%%%%%%%%%%%%%%%%%%%%%%%%%%%%%%%%%%%%%%%%%%%
	In this paper, we briefly reviewed the  domain of existence of the spherically symmetric axion star
	solutions,  considering self-interaction complex scalar field minimally coupled to Einstein’s gravit. The solutions for axion stars at $f_a=1$ are similar to the results obtained for mini-boson stars, suggesting that axion stars can decay to mini-boson stars when $f_a$ is sufficiently large.
	As $f_a$ decreases the self-interaction strengthens the mass-frequency relation changes from the spiral shape to a ``duck-bill" curve. For mini-boson stars, there is only one stable branch, while axion stars at small values of the parameter $f_a$ exhibit a new stable branch in the high-density region. The Newtonian stable branch from the maximum frequency to the first local maximum mass, while the relativistic stable branch from the second $Q \mu^2=M\mu$ to the second mass local maximum. An unstable branch connects the Newtonian and relativistic stable branches, with axion stars having greater compactness under the relativistic stable branch.

	Subsequently, we investigated deformability and the corresponding tidal Love numbers including electric-type and magentic-type.
	Tidal Love numbers are fall into electric-type (even parity) and magnetic-type (odd parity) based on parity.
	We have numerically solved the  ordinary differential equations using the finite element method.
	Expressions for the Love numbers of axion stars were derived through perturbations to the Einstein equations when considering radius distant from the center of axion stars. Following this, six distinct decay constants, denoted as $f_a$, were chosen, and computations were conducted for the quadrupole tidal Love numbers ($\ell=2$) as well as the octupole tidal Love numbers ($\ell=3$).  The results reveal that on the stable branch, whether Newtonian or relativistic, the electric tidal Love number is positive while the magnetic tidal Love number is negative.
	These indicate that even perturbations lead to the deformation which causes gravitational potential that enhance the deformation. But odd perturbations produce spin multipole moments inside axion stars that oppose the external tidal field’s multipole moments.
	And the electric tidal Love numbers are larger than the magnetic tidal Love numbers. With an increase in mass, there is a corresponding decrease in the tidal Love number.
	Those results are simlar with boson stars and Proca stars ~\cite{Cardoso:2017cfl,Herdeiro:2020kba}. However, on the Newtonian stable branch, the electric tidal Love number is significantly larger than the magnetic, whereas on the relativistic branch, the electric tidal Love number is only slightly exceeds the magnetic. Furthermore, the tidal Love numbers on the relativistic stable branch are much smaller than those on the Newtonian stable branch, and as $M/M_{\text{max}}\to1$ on the relativistic branch, the tidal Love numbers tend to approach zero. This intriguing result suggests that at the maximum mass on the relativistic branch, the tidal Love numbers reach their minimum values.
	Due to the more compact configurations of axion stars immersed in external tidal fields, smaller mass and spin multipole moments are produced, thereby resulting in weakened tidal deformability.

	Rotating axion boson stars as the spinning generalisation of static axion boson has been constructed in~\cite{Delgado:2020udb}.
	Extending our study to investigate the tidal Love numbers of rotating axion boson stars and multifield axion boson stars would be interesting.
	Additionally, recent studies indicate the stability of spherically symmetric excited-state boson stars when a self-interaction is introduced in~\cite{Brito:2023fwr,Sanchis-Gual:2021phr}. Therefore, investigating the tidal deformability of ground states and excited states in such self-interacting spherically symmetric configurations is highly meaningful.
	%%%%%%%%%%%%%%%%%%%%%%%%%%%%%%%%%%%%%%%%%%%%%%%%%%%%%%%%%%
	\section*{ACKNOWLEDGEMENTS}
	%%%%%%%%%%%%%%%%%%%%%%%%%%%%%%%%%%%%%%%%%%%%%%%%%%%%%%%%%%
	This work is supported by National Key Research and Development Program of China (Grant No. 2020YFC2201503) and the National Natural Science Foundation of China (Grants No.~12275110 and No.~12047501). Parts of computations were performed on the Shared Memory system at Institute of Computational Physics and Complex Systems in Lanzhou University. 	
	%%%%%%%%%%%%%%%%%%%%%%%%%%%%%%%%%%%%%%%%%%%%%%%%%%%%%%%%%%%%%%%%%%%%%%%%%%%%%%
	
	%%%%%%%%%%%%%%%%%%%%%%%%%%%%%%%%%%%%%%%%%%%%%%%%%%%%%%%%%%%%%%%%%%%%%%%%%%%%%%

\end{document}